\documentclass[prl, twocolumn, superscriptaddress]{revtex4-1}
\usepackage{amsmath}
\usepackage[english]{babel}
\usepackage[pdftex]{graphicx}
\usepackage{color}
\usepackage{ulem}
\usepackage{comment}

\usepackage[sort&compress]{natbib}
\usepackage{hyperref}
\graphicspath{{images/}{./}}

\usepackage{amsmath}
\usepackage{amssymb}
\usepackage{mathtools}
\usepackage{braket}

\renewcommand{\vec}[1]{\ensuremath{\boldsymbol{#1}}}

\begin{document}

\title{Twofold topological phase transitions induced by third-nearest-neighbor hoppings in 1D chains}

\author{Yonatan Betancur-Ocampo}
\email{ybetancur@fisica.unam.mx}
\affiliation{Instituto de F\'isica, Universidad Nacional Aut\'onoma de M\'exico, Ciudad de México, M\'exico}

\author{B. Manjarrez-Monta\~nez}
\affiliation{Instituto de Ciencias F\'isicas, Universidad Nacional Aut\'onoma de M\'exico, Cuernavaca, M\'exico}

\author{A. M. Mart\'inez-Arg\"uello}
\affiliation{Instituto de F\'isica, Benemérita Universidad Autónoma de Puebla, C. P. 72570 Puebla, Puebla, Mexico}

\author{Rafael A. M\'endez-S\'anchez}
\email{mendez@icf.unam.mx}
\affiliation{Instituto de Ciencias F\'isicas, Universidad Nacional Aut\'onoma de M\'exico, Cuernavaca, M\'exico}

\begin{abstract}
Strong long-range hoppings up to third nearest neighbors may induce a topological phase transition in one-dimensional chains. Unlike the Su-Schrieffer-Heeger model, this transition from trivial to topological phase occurs with the emergence of a pseudospin valley structure and a twofold nontrivial topological phase. Within a tight-binding approach, these topological phases are analyzed in detail and it is shown that the low-energy excitations follow a modified Dirac equation, in which the dynamics of particles with positive and negative mass occur differently. An experimental realization in a one-dimensional elastic chain, where it is feasible to tune directly the third-nearest-neighbor hoppings, is proposed. 
\end{abstract}

\maketitle

{\it Introduction.} 
Topological phase transitions are one of the most investigated topics in condensed matter physics nowadays~\cite{Pal2016,Verbin2013,Rufo2019,Pieczarka2021,Zhang2022,Chen2022,Thatcher2022}. 
With the increasing interest in topological insulators, the discovery of novel and exotic states of matter continues~\cite{Drost2017,Li2019,Liu2013,Pernet2022,Xie2019,Meier2016,Wang2022}. 
In many cases, the experimental realization of outstanding phenomena predicted by some models has been inaccessible to date in condensed-matter laboratories. In this regard, artificial systems have been used as ancillary tools to test the predictions that emerge from those models~\cite{Yang2016,Wu2021,Zhang2015,Mukherjee2015,Huda2020,Belopolski2017,Freeney2022,Tang2022,Stegmann2017,Pernet2022}. 
In addition, they also allow to explore regimes almost impossible to reach in synthesized materials; a concrete example is the long-range interaction \cite{Li2019,PerezGonzalez2018,Zhang2022,Tang2022}. Usually, the hopping parameter among atomic sites decreases with its relative separation \cite{Stegmann2017,Bellec2013,Martinez-ArguelloEtAl2022,RamirezRamirez2020}. In artificial systems, however, those hopping parameters may be engineered {\sl ad hoc}~\cite{Lopez-ToledoEtAl2021}. 
This versatility opens the door to tackle abundant novel physics hardly reachable in to-date condensed matter experiments. In artificial elastic systems, furthermore, high-order couplings can be engineered as well as nearest-neighbor couplings, a difficult task to reach in optical and acoustic systems.

The most successful description of topological insulators is provided by the Su-Schrieffer-Heeger (SSH) model~\cite{Su1979,Heeger1981,Heeger1988,Fradkin1983}. This model predicts outstanding phenomena such as the creation of solitons~\cite{Su1979}, the conductivity in polymers~\cite{Heeger1988}, and the simplest explanation of a topological phase transition in trans-polyacetylene~\cite{Asboth2016}. 
Up to now, there are multiple experimental realizations of the SSH model in artificial systems ~\cite{Chen2022a,Meier2016,Li2018,CaceresAravena2022,Thatcher2022,Kiczynski2022,Coutant2021}, as well as extensions of the SSH model that depict unusual effects such as robust edge states \cite{CaceresAravena2022}, non-hermitian skin-effect due to spin-orbit interactions \cite{SESSH}, Dirac states \cite{DSSSH}, anomalous diffusion in disordered and nonlinear chains \cite{Manda2023}, and gap solitons \cite{Pernet2022}. 
The coupling between two SSH chains has also been explored recently using the generalized Rice-Mele model \cite{McCann}.

In this work, an effective tight-binding model depicts a chain with first- and third-nearest-neighbor hoppings, which resembles the cis-polyacetylene molecule of organic chemistry. 
A twofold topological phase transition is induced by modulating the third-nearest-neighbor hoppings in regimes outside the typical values of the hopping scaling rule of carbon bonds in molecular systems. Our starting point is a tight-binding model of four atoms in the unit cell. 
From the continuum approximation, we found that the reduced effective Hamiltonian for two-level systems differs from the SSH model since it describes two topological phases and not only one. 
The chain considered here is identical to two-coupled SSH chains, or well a ribbon of a 2D SSH model~\cite{LiuWakabayashi,LiuDengWakabayashi}. Effects due to the folding of the first Brillouin zone and band structure are expected using a 4-band formalism. 
The emergence of a pseudo-spin valley structure leads to the possibility of realizing valleytronics in one-dimensional systems. This model also includes a mass conjugation symmetry breaking, where negative and positive mass particles possess different dynamics. The edge states in a finite version of the chain are also studied. An experimental realization that captures the twofold topological phase is proposed for the corresponding chain in an elastic structure.

\begin{figure*}[t!]
    \centering
    \begin{tabular}{c}
       \includegraphics[scale=0.5]{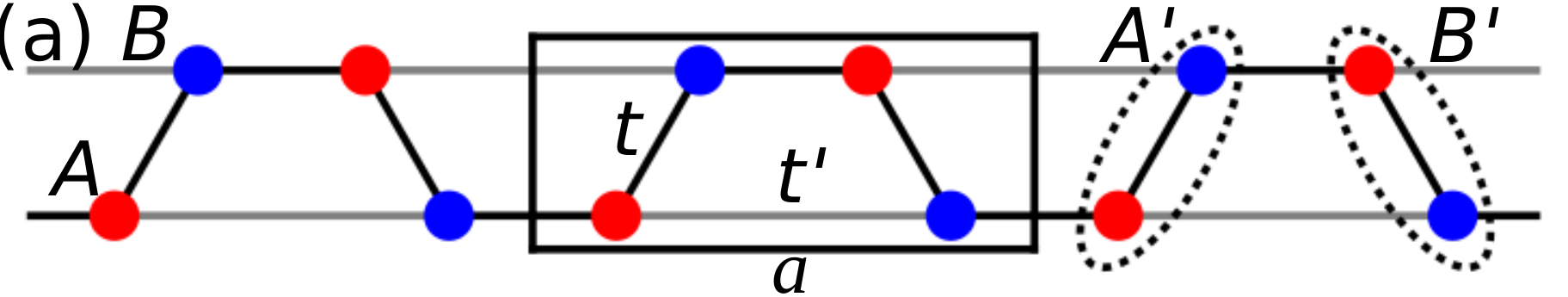}  
    \end{tabular}
    \begin{tabular}{c c c c}
     \includegraphics[scale=0.275]{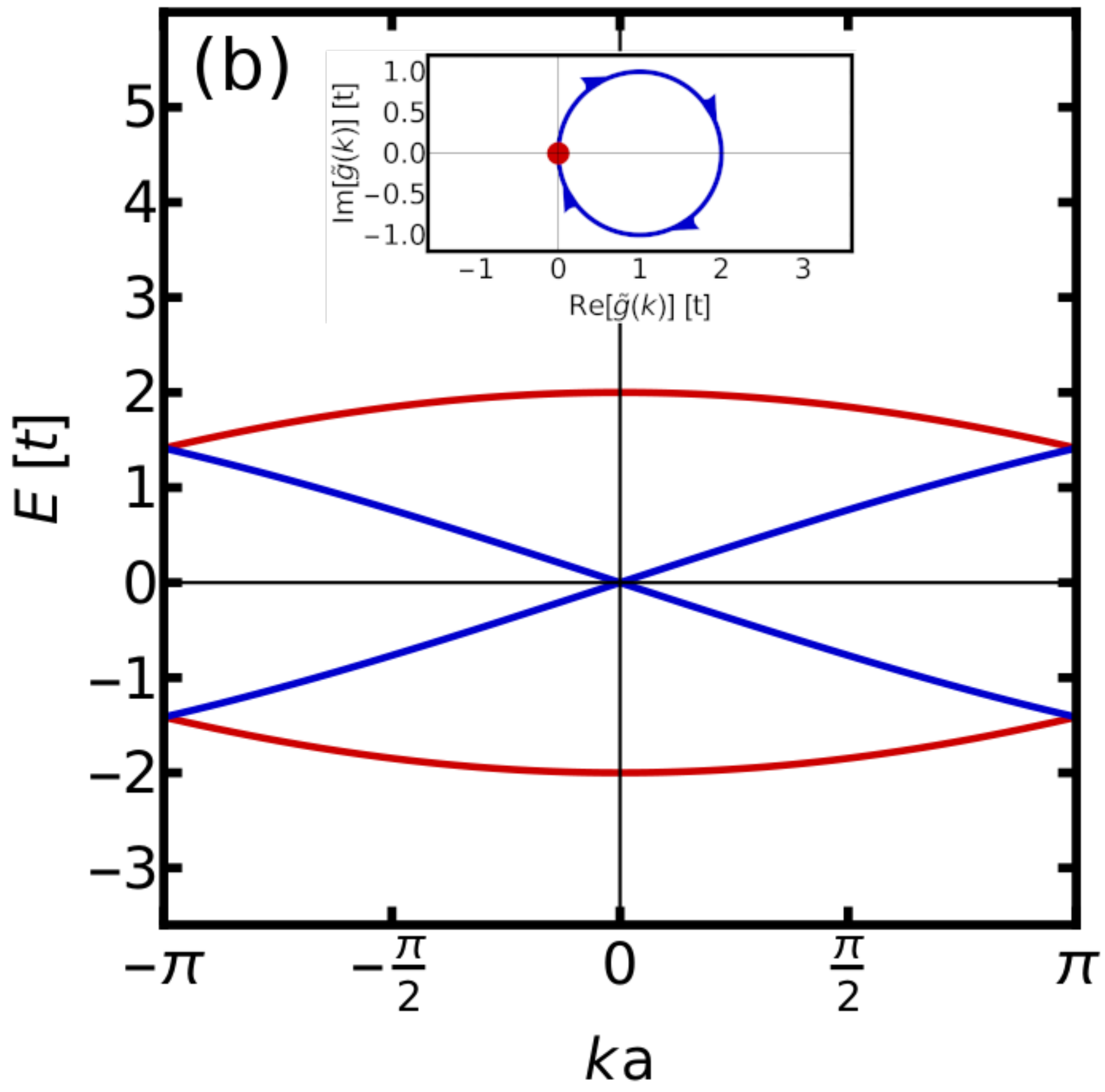} &  
     \includegraphics[scale=0.275]{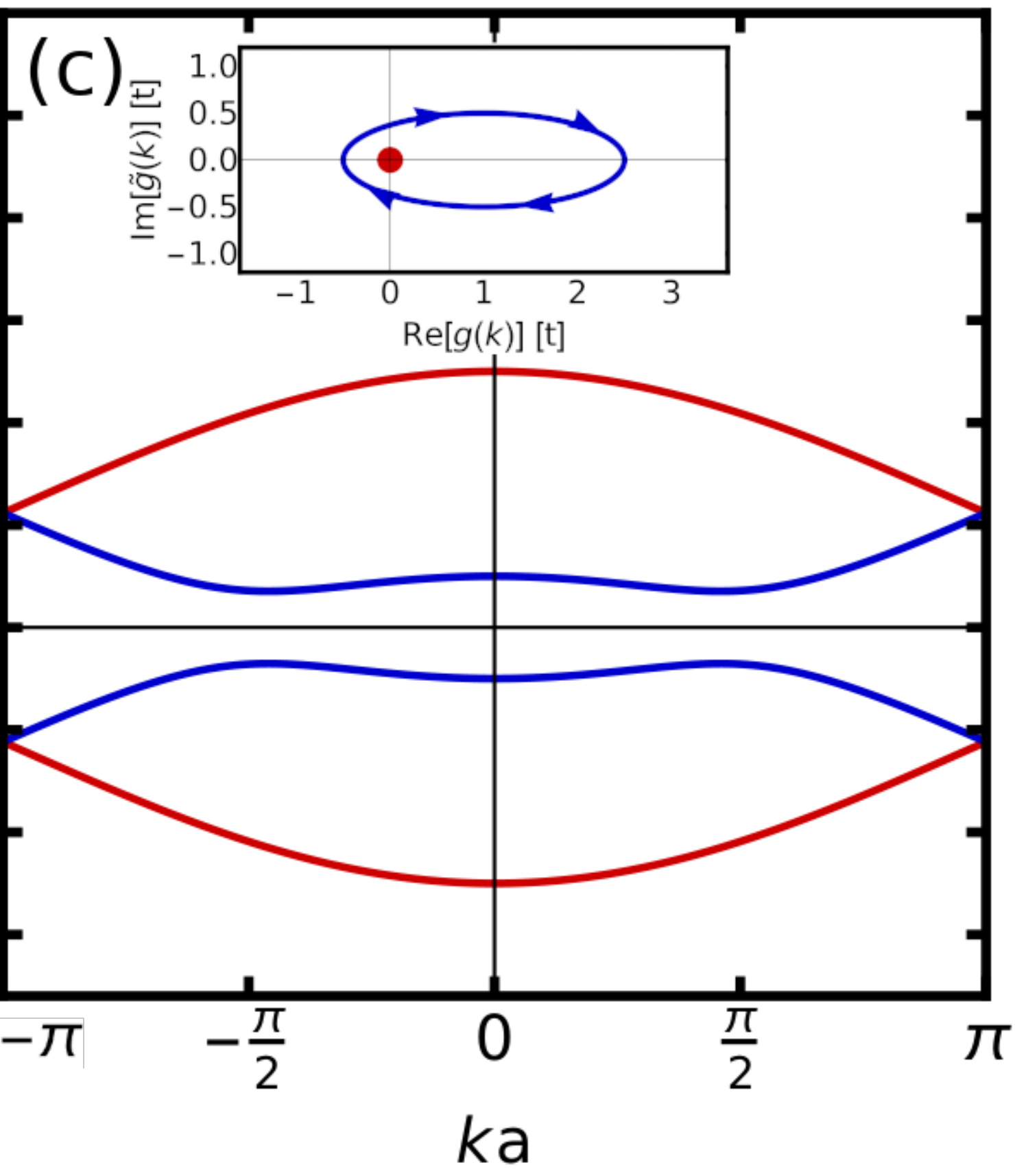}
      \includegraphics[scale=0.275]{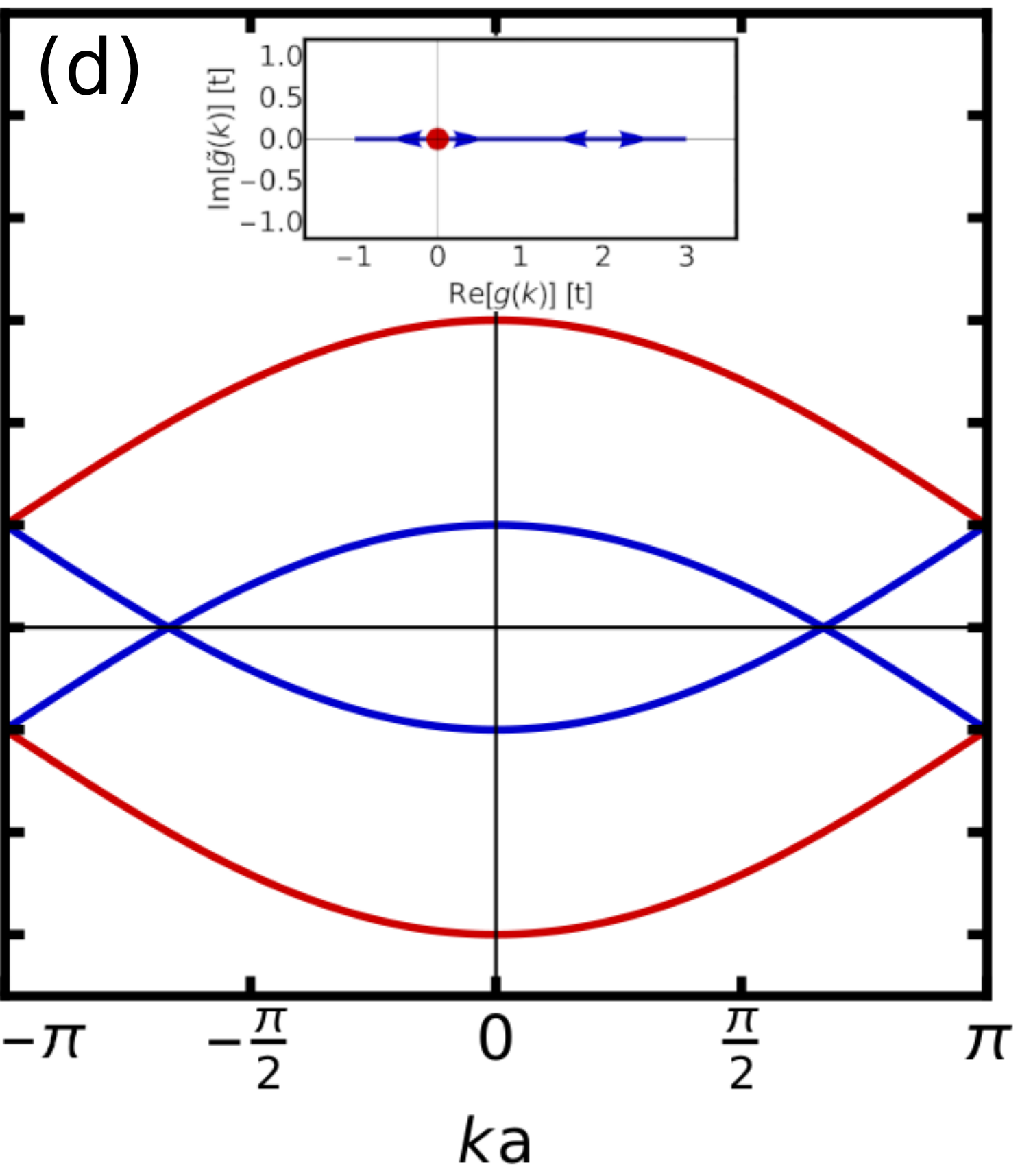} &
      \includegraphics[scale=0.275]{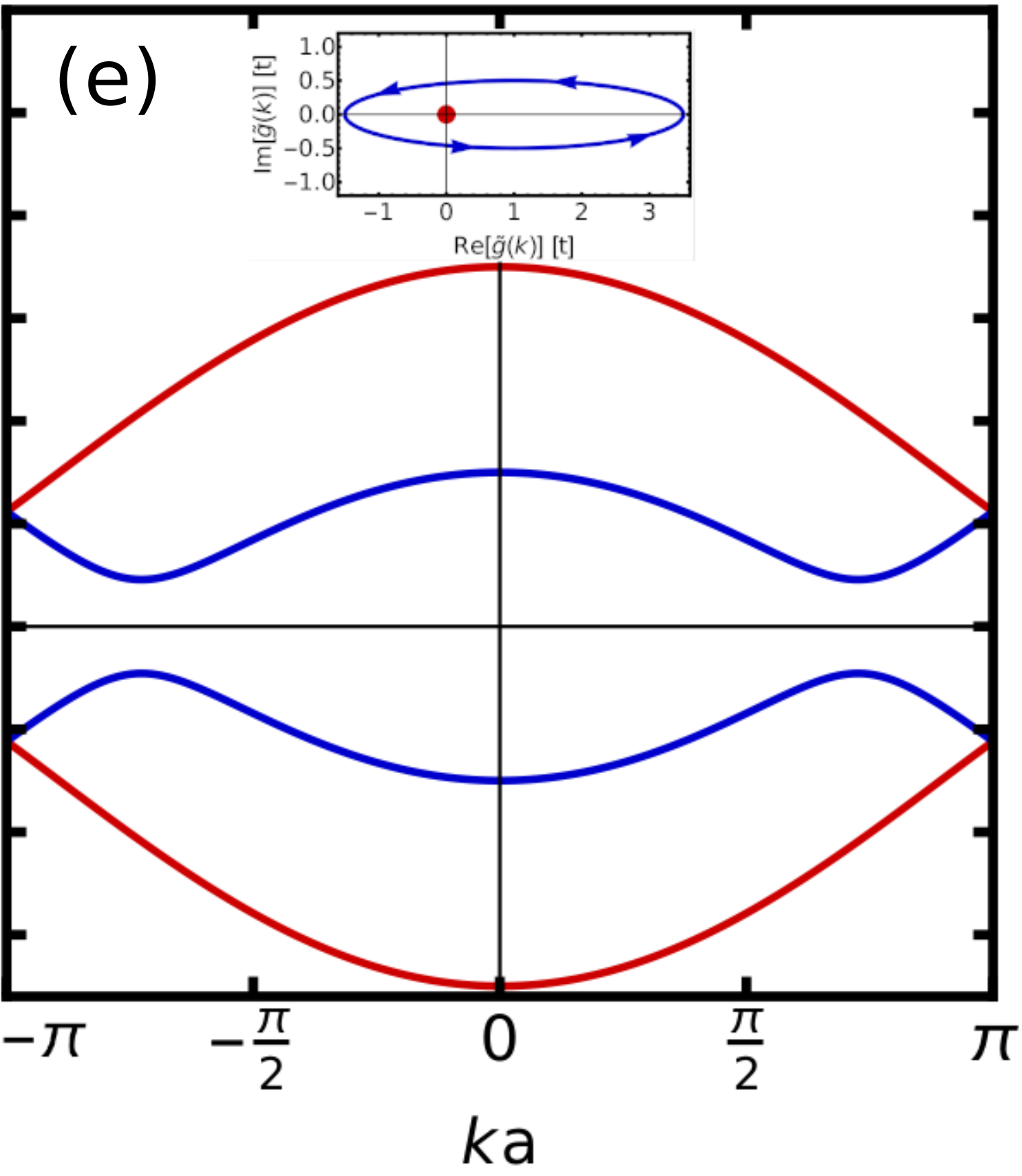}
    \end{tabular}
    \caption{(a) The one-dimensional lattice with first and third-nearest-neighbors. The rectangle indicates the unit cell with sites $A$ and $B$ for the sublattices, which are discriminated by the $\sigma_j$ Pauli matrices. The dimers $A'$ and $B'$ can also be seen as the sites of other sublattices, associated with the $\tau_j$ Pauli matrices. The parameters $t$ and $t'$ correspond to the hoppings to first and third-nearest-neighbors, respectively. The lattice constant is $a$. The red (blue) bands in (b)-(e) correspond to upper (lower) energy. The insets show the pseudo-spin trajectory in the Bloch plane with a topological charge (which is represented by a red disk) and winding number as a function of $t'$ with $t = 1$. (b) Band structure of the armchair edge of graphene with first-neighbor hoppings only. The pseudo-spin trajectory passes just on the topological charge and the winding number is undefined. (c) Turning on the third-nearest-neighbors ($t' = 0.5$), a band gap is opened, and the winding number is $w = -1$. The topological charge is found inside the loop. (d) When $t' = t$, a closing of the gap is observed and the winding number becomes undefined. (e) For $t' > t$, the band gap is reopened and the direction of the loop is inverted. The winding number is $w = 1$.}
    \label{fig:Polyacetylene}
\end{figure*}

{\it Tight-binding model of a one-dimensional chain with third-nearest-neighbor hoppings.} 
In Fig.~\ref{fig:Polyacetylene}, a one-dimensional chain with hoppings to first and third nearest neighbors, excluding the second ones, is proposed. 
In contradistinction with the SSH model two hopping parameters $t$ and $t'$, that correspond to couplings to first and third nearest neighbors, respectively, will be used. 
Using a tight-binding model on a Fourier basis, the Bloch Hamiltonian is given by
\begin{equation}\label{Hpol}
    H
    (k)=\left(\begin{array}{c c c c}
        E_0  & g^*(k) & 0 & h^*(k) \\
         g(k)  & E_0 & h(k) & 0 \\
         0  & h^*(k) & E_0 & g^*(k)\\
         h(k) & 0 & g(k) & E_0
    \end{array}\right),
\end{equation}
where $E_0$ is the site energy, $g(k)= t\exp(\mathrm{i} ka/6)$, $h(k)= h_x(k) + \mathrm{i} h_y(k) = t\exp(-\mathrm{i} ka/3) + t'\exp(2 \mathrm{i} ka/3)$, $k$ is the wave vector, and $a$ is the lattice constant. 
Here, it is assumed that the sites possess identical orbitals and therefore the site energies $E_0(=0)$ are the same. 
In order to define the pseudospin structure of the chain in Fig.~\ref{fig:Polyacetylene}(a), the Hamiltonian in Eq.~\eqref{Hpol} can be expressed as
\begin{equation}\label{Hpol2}
    H(k)= \tau_0\otimes\vec{\sigma}\cdot\vec{g}(k) + \tau_x\otimes\vec{\sigma}\cdot\vec{h}(k),
\end{equation}
where $\vec{\sigma} = (\sigma_x, \sigma_y)$ are the Pauli matrices that discriminate the sublattices $A$ and $B$, and $\tau_0$ and $\tau_x$ are also Pauli matrices acting on the pairs $A'$ and $B'$, see Fig.~\ref{fig:Polyacetylene}(a). 
The pseudospin vectors are defined as $\vec{g}(k)=(g_x(k),g_y(k))$ and $\vec{h}(k)=(h_x(k),h_y(k))$. From the Hamiltonian in Eq.~\eqref{Hpol2}, the energy bands and spinors are 
\begin{equation}\label{eb0}
    E_{s,\nu}(k)= s|g(k)+\nu h(k)| \,\, \textrm{and} \,\, |\psi^\nu_s\rangle = \frac{1}{2}(1,s\textrm{e}^{\mathrm{i}\phi_\nu},\nu,s\nu\textrm{e}^{\mathrm{i}\phi_\nu}),
\end{equation}
respectively, where the pseudospin angle is 
\begin{equation}
    \phi_\nu = \arctan\left(\frac{\textrm{Im}[g(k)+\nu h(k)]}{\textrm{Re}[g(k)+ \nu h(k)]}\right),
    \label{eq:pseudospin}
\end{equation}
and the band index $s = 1$ ($= -1$) for the conduction (valence) band, while the index $\nu = -1$ ($=1$) indicates the lower (upper) band. 

Notice that the model in Eqs.~\eqref{Hpol2} and~\eqref{eb0} describes a different topological phase transition than that of the SSH model~\cite{Asboth2016}. In fact, this new model establishes different regimes depending on the third-nearest-neighbor hopping $t'$ by letting fixed the first one to $t = 1$. When $t' = 0$, the band structure of the chain is obtained, which is the typical case of energy bands for the edge of an armchair graphene nanoribbon (cis-polyacetylene) to first nearest neighbors. This is observed in Fig. \ref{fig:Polyacetylene}(b) in red and blue curves which correspond to the indices $\nu = -1$ and $1$, respectively (see Eq.~\eqref{eb0}). In the regime $t' < 1$, see Fig.~\ref{fig:Polyacetylene}(c), the semimetallic phase is destroyed by inducing a gap and two minima in the conduction band appear. Such a feature shows the formation of a new valley pseudospin structure, which is linked to the time-reversal symmetry operation. In the case $t' = 1$, as shown in Fig.~\ref{fig:Polyacetylene}(d), one-dimensional Dirac cones in each valley emerge. It is worth noting that the Fermi velocity for excitations traveling from left to right  differs from that for excitations traveling from right to left. A reopening of the band gap is observed in the range $t' > 1$. This process of closing and reopening of the band gap is an indicative of a topological phase transition. The phases shown in Fig. \ref{fig:Polyacetylene}(b)-(e) offer the possibility to study valley-dependent transport properties in one-dimensional chains.

In what follows, the topological phase transition is investigated by proposing a more simplified model than the one given in Eq.~\eqref{Hpol2}. The simplification consists in reducing the $4 \times 4$ Hamiltonian to a $2 \times 2$ one. For the dispersion relation in Eq.~\eqref{eb0}, one can identify that $\nu = -1$ is the dispersion relation for the lower bands, which is the reminiscence of anti-bonding in molecules. 

Any symmetric two-level system has a Hamiltonian given by
\begin{equation}
    H_\textrm{2level}(k) = \left(\begin{array}{cc}
        0 & \tilde{g}^*(k) \\
        \tilde{g}(k) & 0
    \end{array}\right),
    \label{eq:H2level}
\end{equation}
whose eigenenergies and eigenfunctions are 
\begin{equation}\label{eb}
    E_{s}(k)= s|\tilde{g}(k)| \,\, \textrm{and} \,\, |\psi_s\rangle = \frac{1}{\sqrt{2}}(1,s\textrm{e}^{\mathrm{i}\phi_-}).
\end{equation}
It is straightforward to link the function $\tilde{g}(k)$ with $g(k)$ and $h(k)$ as $\tilde{g}(k) = g(k) - h(k)$. 
The pseudo-spin angle $\phi_-$ has an identical expression as in Eq.~\eqref{eq:pseudospin} setting $\nu=-1$. 
In order to describe the topological phase transition in Fig.~\ref{fig:Polyacetylene}, the function $\tilde{g}(k)$ is expanded around the center of the First Brillouin zone up to second order in $k$. 
That is,
\begin{equation}
    \tilde{g}(k) \approx -t' + \left(2t'-\frac{3}{2}t\right)\mathrm{i}ka + \left(2t'+\frac{3}{8}t\right)(ka)^2.
\end{equation}
This allows us to build an effective model in the continuum approximation given by
\begin{equation}\label{Heff}
    H_\textrm{eff}(k) = \left(-\Delta + \frac{k^2}{2\mu}\right)\sigma_x + vk\sigma_y,
\end{equation}
where the effective parameters are related to the hoppings $t$ and $t'$ as  $\Delta = t'$, $1/(2\mu) = (2t' + 3t/8)a^2$ and $v = (2t' - 3t/2)a$. 
The effective Hamiltonian in Eq.~\eqref{Heff} is identical to the one-dimensional version of the modified Dirac Hamiltonian \cite{SHUNQING2011,Shen2017} and describes the topological phase transition shown in Fig. \ref{fig:Polyacetylene}. 
The squared term of the wave vector in Eq.~\eqref{Heff} breaks the positive and negative mass symmetry of the conventional Dirac equation. It is important to mention that the Hamiltonian in Eq.~\eqref{Heff} differs the SSH one because the latter describes in an identical way the dynamics of particles with positive and negative mass.

Another remarkable feature with respect to the SSH model is the pseudo-spin valley structure observed in Fig.~\ref{fig:Polyacetylene}(b)-(e). 
Since the unit cell of the chain possesses four sites, this causes an energy band folding, where two valleys inside the first Brillouin zone emerge, as shown in Fig.~\ref{fig:Polyacetylene}. 
This pseudo-spin valley structure is absent in conventional SSH chains. 
The third nearest neighbor cis-polyacetylene lattice can be viewed as two coupled SSH chains instead of a single one, where a two-fold topological phase transition appears. 
To recognize the type of low energy excitation in Fig.~\ref{fig:Polyacetylene}, it is necessary to obtain the effective Hamiltonian around the valleys. We find that, at the valleys $K^+$ and $K^-$, which are located at the points $K^\pm = \pm \sqrt{2\mu(\Delta+\mu v^2)}$, the effective Dirac-like Hamiltonian is
\begin{equation}
    H^\pm_D(q) = (\pm v'\sigma_x + v\sigma_y)q + \mu v(v\sigma_x \pm v'\sigma_y),
\end{equation}
where $q = k-K^\pm$ is the wave vector near the valley $K^\pm$ and the velocity $v' = \sqrt{2(\Delta + \mu v^2)/\mu}$. This feature is similar to pseudo-spin valley structure of graphene and the complete $4 \times 4$  Hamiltonian that depicts both valleys is given by $H=\textrm{diag}[H^+_D(q),H^-_D(q)]$. The dispersion relation around $K^\pm$ points is $E^\pm_s = s\sqrt{(\mu v^2 \pm qv')^2 + (qv \pm \mu v^2)^2}$, which is anisotropic because the group velocities for particles from $K^\pm$ are given by
\begin{equation}
    v^\pm_g(q) = s\frac{qv'^2+\mu v^2v' \pm(q v^2 + \mu vv'^2)}{\sqrt{(\mu v^2 \pm qv')^2 + (qv \pm \mu v^2)^2}}.
\end{equation}
This indicates that the propagation wave for one of the valleys is slower than the other valley. Setting $q = 0$, we have the simple expression $v^\pm_g(0) = sv'(v \pm v)/\sqrt{v^2+v'^2}$, which are the velocities at the minimum (maximum) of the conduction (valence) band. Therefore, such a third nearest neighbor chain offers the possibility to manipulate the valley degree of freedom in the quantum transport of one-dimensional chains.

{\it Topological characterization of the 1D chain.} To identify the topological phases described above, we resort to the bulk-boundary correspondence with the calculation of the winding number $w$, which is related to the Berry phase $\beta_s = s\pi w$, where $s$ is the band index. We look for regimes of $t$ and $t'$ where $\beta_s$ is invariant. This can be illustrated by taking into account the trajectory of the pseudo-spin vector in the Bloch plane and identifying the loops which enclose the origin, where the topological charge is located. The Berry curvature for any symmetric band two-level system is $\vec{B}_s = s\hat{\tilde{\vec{g}}}(k)/(2|\tilde{\vec{g}}(k)|^2)$ \cite{BERRY2017,Zak1989,Asboth2016}. Therefore, it always exists a divergence when the conduction and valence bands touch one to each other \cite{BERRY2017,Zak1989,Asboth2016}. When the pseudo-spin vector loop encloses a topological charge, the winding number is different from zero. The insets in Fig. \ref{fig:Polyacetylene}(b)-(e) help to recognize the different topological phases. When $t' = 0$, which is the case of cis-polyacetylene without third-nearest-neighbor hoppings, a semimetallic phase is obtained. Turning on the hopping parameter $t'$ is sufficient for obtaining topological phases. As in the SSH model, the transition has also a closing and reopening band gap. 
However, the model in Eq.~\eqref{Hpol} shows two topological phases at this transition with opposite winding numbers $w = -1$ and $w = 1$ for the regimes $t' < t$ and $t' > t$, respectively. 
These phases can be understood by the change of orientation of the loop.
\begin{figure*}
    \centering
     \begin{tabular}{cccc}
     (a) \qquad \qquad \qquad \qquad \qquad & (b)\qquad \qquad \qquad \qquad \qquad & (c) \qquad \qquad \qquad \qquad \qquad & (d)\qquad \qquad \qquad \qquad \qquad \\
     \includegraphics[scale=0.24]{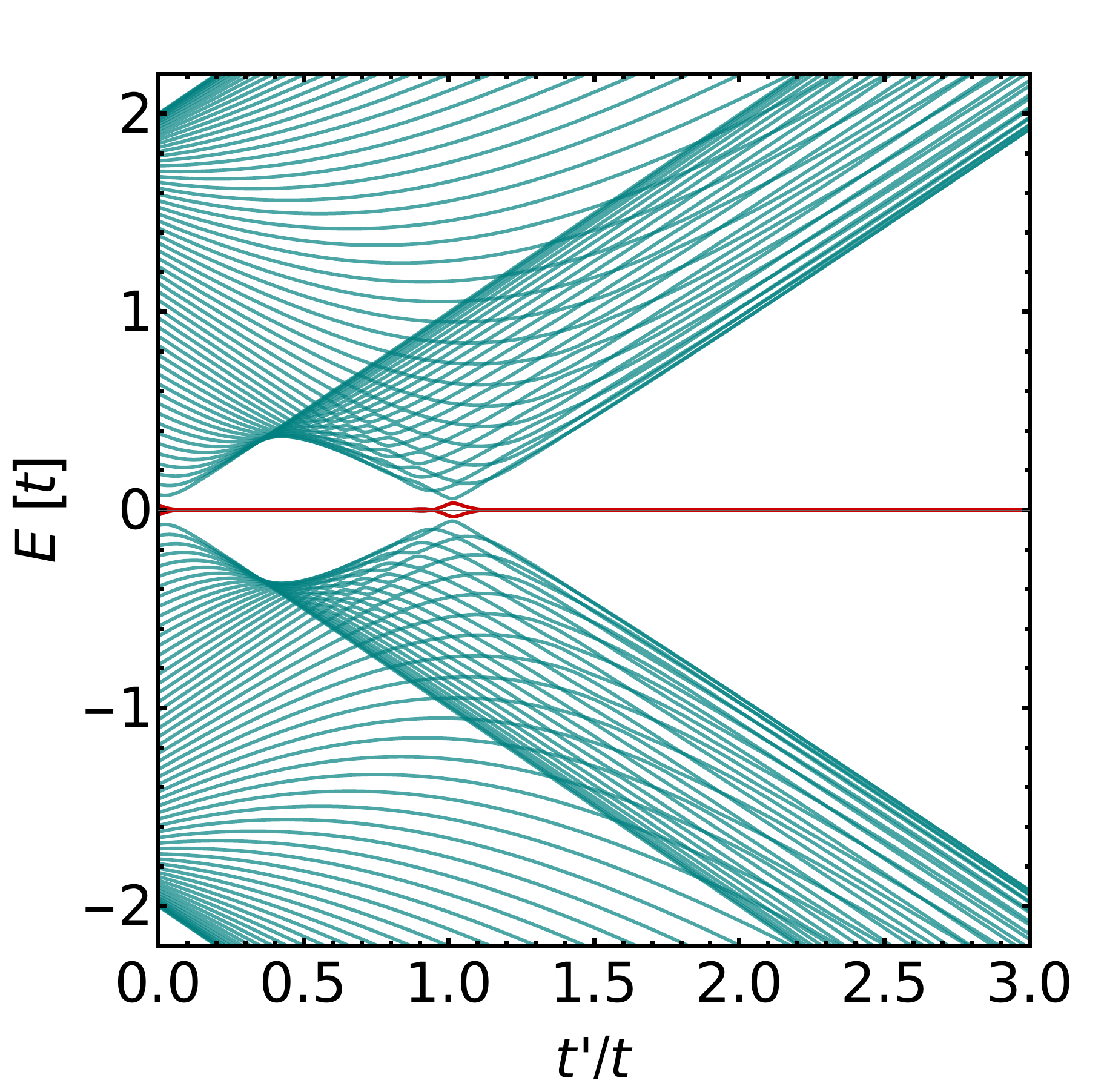} &  
     \includegraphics[scale=0.25]{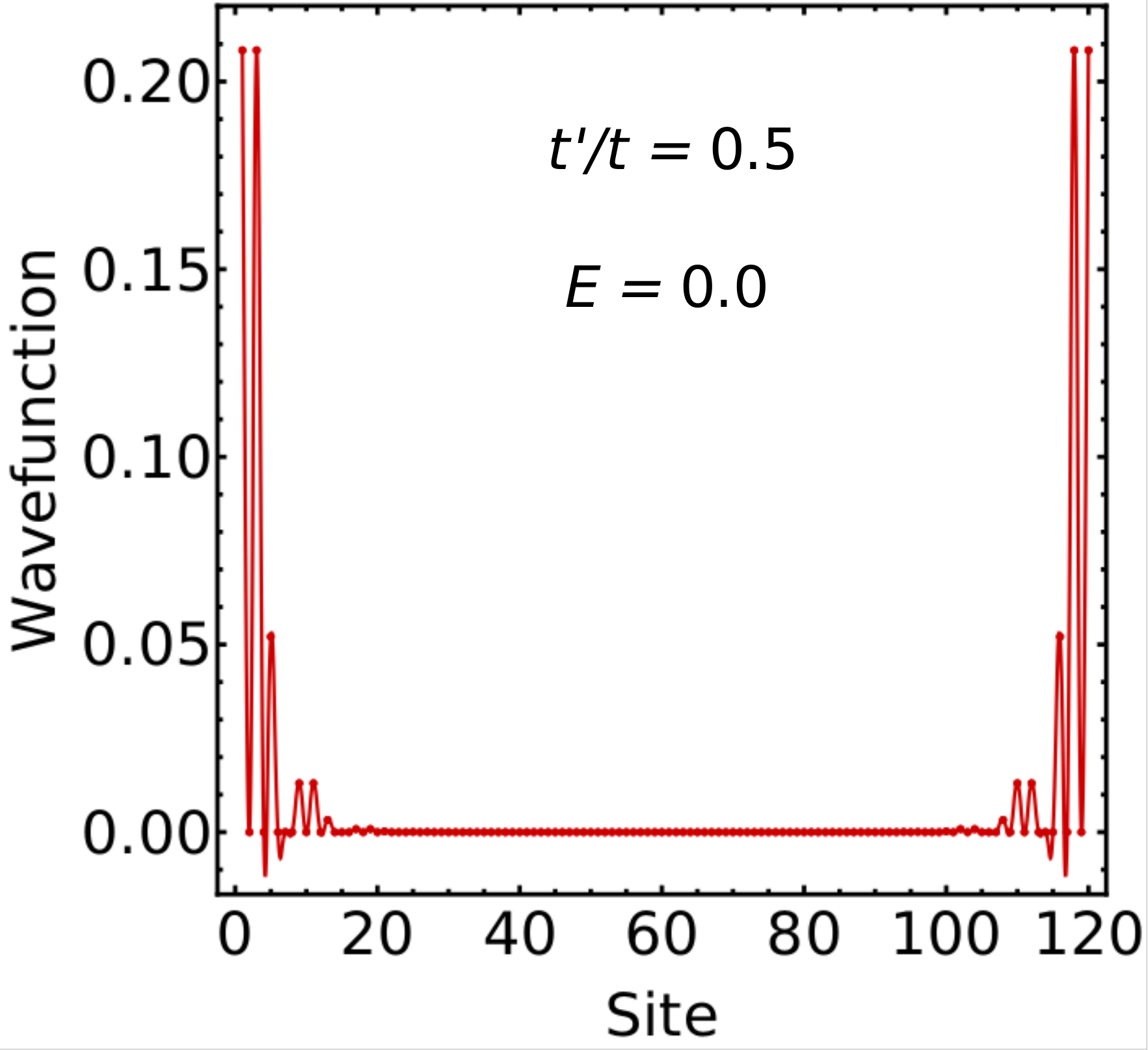}&
    \includegraphics[scale=0.25]{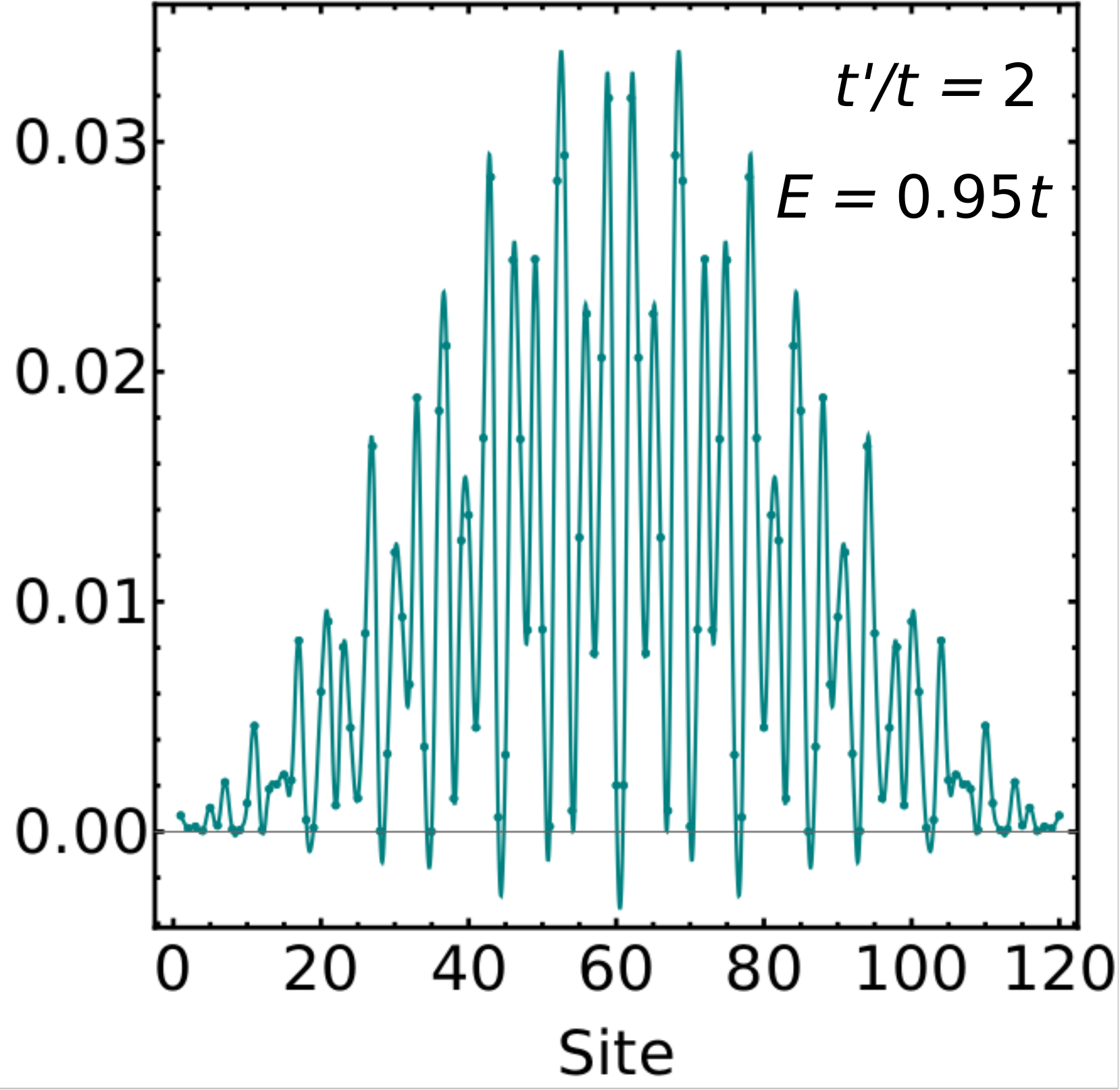}&
    \includegraphics[scale=0.25]{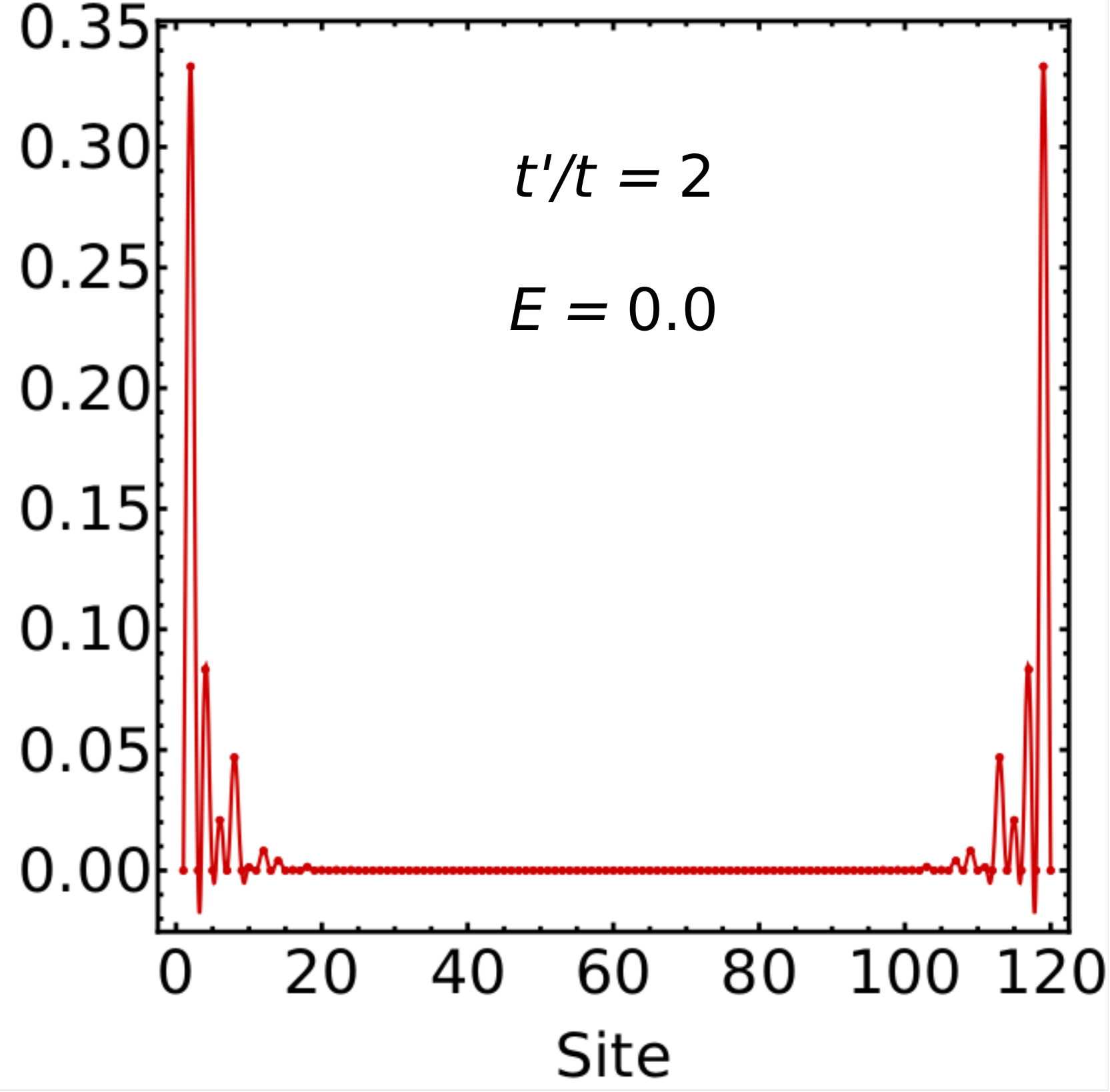}
    \end{tabular}
    \caption{(a) Energy bands of a finite third-nearest-neighbors cis-polyacetylene as a function of the hopping parameters ratio $t'/t$. The turquoise curves corresponds to bands for bulk states, while red ones at the center for edge states. (b)-(d) wave function as a function of the site $j$. Depending on the energy and hopping ratio $t'/t$, we can obtain edge or bulk states. In (b) we set $E = 0$ and $t'/t = 0.5$ and the wave is localized at the edges, which is unaffected changing the value of $t'/t$. An similar behavior is obtained in (d) with $t'/t = 2$. In contrast, if we set $E = 0.95 t$ and $t'/t = 2$ bulk states are observed in (c).}
    \label{fig:Finite}
\end{figure*}

\label{Section:Artificial}

\begin{figure}
    \centering
     \begin{tabular}{c c}
     \includegraphics[scale=0.35]{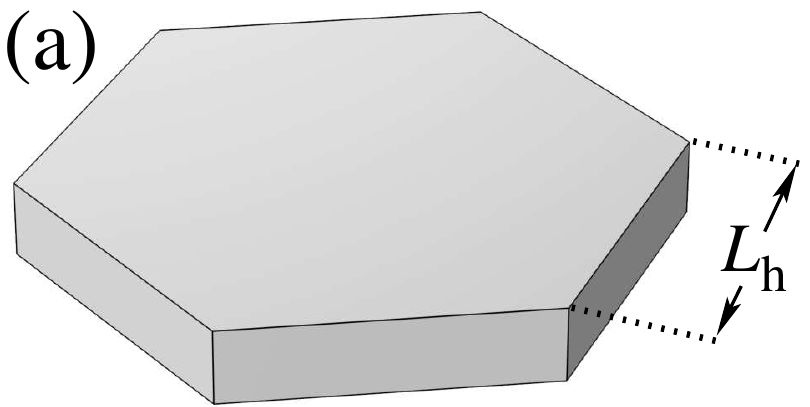} &  
     \includegraphics[scale=0.35]{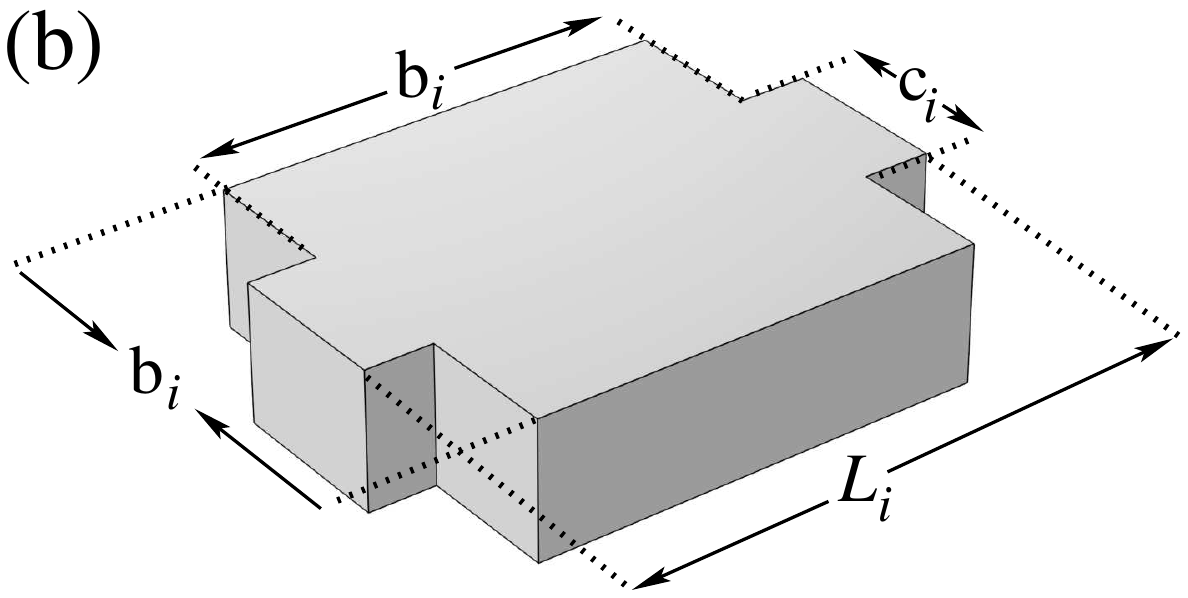}
    \end{tabular} \\
    \includegraphics[scale=0.35]{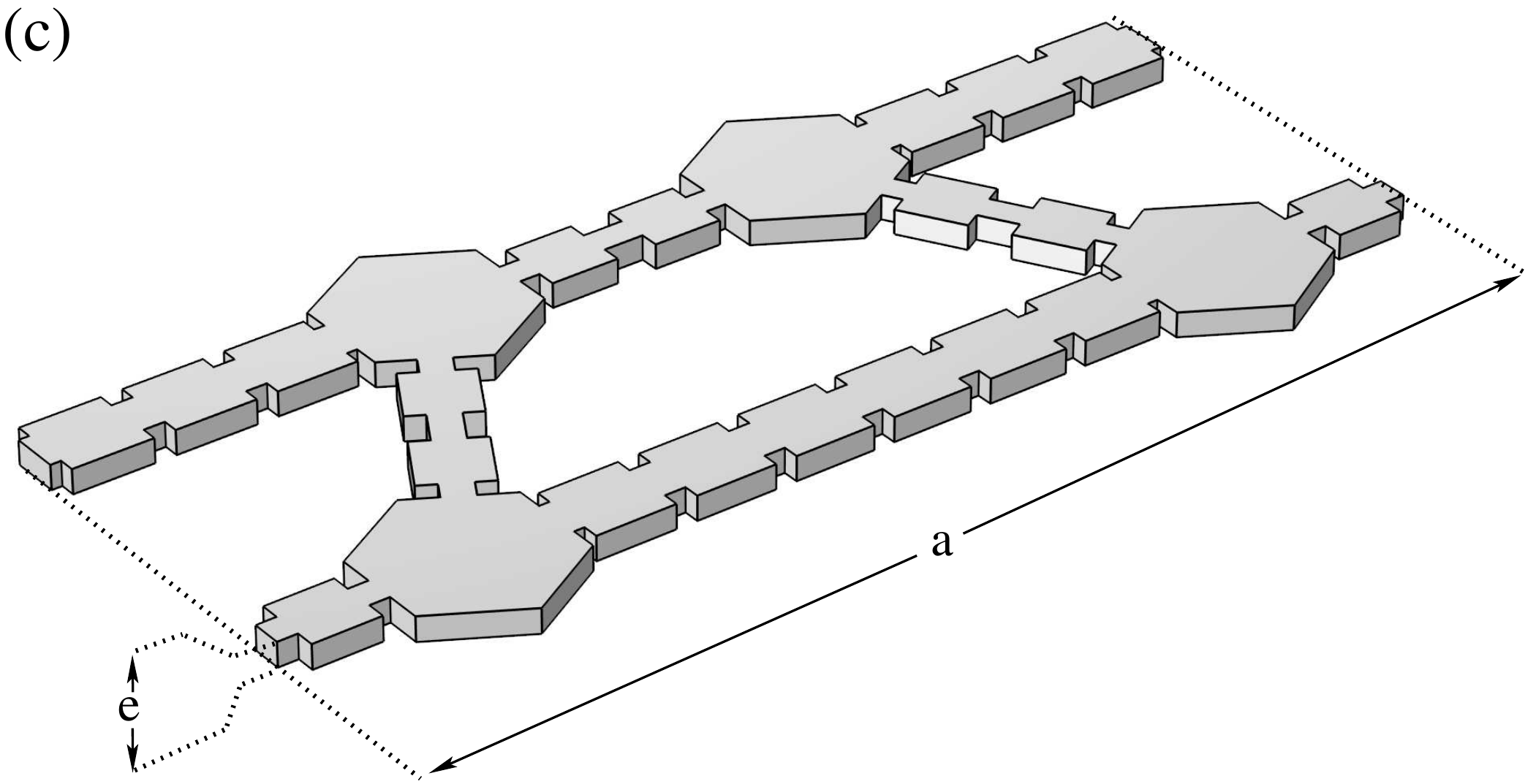}
    \caption{(a) Atomic site composed of a hexagonal-shaped resonator of length $L_h$. The bond is a finite phononic crystal (FPhC) whose unit cell, given in (b), is composed of a square-shaped aluminum plate of length $b_i$ and two smaller plates with length $c_i \times (L_i-b_i)$, one on each side of the square with $i=1,3$ for first and third nearest neighbors, respectively. The number of cells in the FPhC and the width of the smaller plates $c_i$ control the couplings to first and third nearest neighbors. (c) Elastic artificial 1D chain composed of aluminum plate with thickness $e$ and length $a$ coupled through first and third nearest neighbors by two bonds at an angle of $60^\circ$ between them. The geometrical parameters are $e=6.35$~mm, $L_h= 27$~mm, $L_1= 22$~mm, $b_1 = 17$~mm, $c_1 = 2$~mm, $L_3 = 22.4608$~mm, $b_3 = 17$~mm, and $c_3 = 5, 7, 9, 10$~mm.}
    \label{fig:CREW}
\end{figure}
In order to observe whether the topological phases predicted by the model in Fig. \ref{fig:Polyacetylene}(b)-(e) appear also in finite cis-polyacetylene, we consider the tight-binding approach applied to a finite chain. This Hamiltonian consists of a sub-matrix $h_\textrm{cp}$ repeated $N$ times in a block-diagonal form, which is given by
\begin{equation}
    h_\textrm{cp}(t,t') = \left(\begin{array}{cccc}
         E_0 & t & 0 & t'  \\
         t & E_0 & t & 0  \\
         0 & t & E_0 & t  \\
         t' & 0 & t & E_0  
    \end{array}\right).
\end{equation}

\noindent Each block-diagonal matrix is coupled with the nondiagonal submatrix $C$

\begin{equation}
    C(t,t') = \left(\begin{array}{cccc}
         0 & 0 & 0 & 0  \\
         0 & 0 & 0 & 0  \\
         0 & t' & 0 & 0  \\
         t & 0 & 0 & 0  
    \end{array}\right). 
\end{equation}

\noindent Therefore, the Hamiltonian for the finite chain has a Toeplitz matrix form

\begin{align}
\label{eq:Nssh}
    H_\textrm{fc}(t,t') = \qquad \qquad \qquad \qquad \qquad \qquad \qquad \qquad \qquad &\nonumber\\
    \left(\begin{array}{c c c c c}
        h_\textrm{cp}(t,t')  & C(t,t') & 0 & \cdots & 0  \\
         C^\dagger(t,t') & h_\textrm{cp}(t,t') & C(t,t')  & 0 & \cdot\\
         0 & C^\dagger(t,t') & \cdot & \cdot & \cdot \\
         \vdots & \vdots &  \vdots & \vdots & C(t,t') \\
         0 & 0 & 0 & C^\dagger(t,t') & h_\textrm{cp}(t,t')
    \end{array}\right).&
\end{align}

\noindent In total, thirty unit cells are taken into account to obtain the energy bands in Fig. \ref{fig:Finite}(a), which corresponds to $N = 30$ repetitions of the block-matrix $h_\textrm{cp}$. 
This band structure is shown as a function of the hopping parameters ratio $t'/t$. 
We highlighted the energy bands in red to indicate the bands of edge states, which are twofold topologically by examining the whole range of $t'/t$. 
The eigenstates of the Hamiltonian, as shown in Fig. \ref{fig:Finite}(b) and (d), evidence that the wave function is localized in the edges of the chain with an amplitude decaying exponentially to the advance  (1)across the sites. 
This localization is unaffected by the variation of the ratio $t'/t$, which has a robustness characteristic of topological insulators. 
If we set some energy value for the turquoise bands in Fig. \ref{fig:Finite}, the bulk state will appear distributed along the chain, as shown in Fig. \ref{fig:Finite}(c).

\begin{figure*}
\centering
\includegraphics[width=\textwidth]{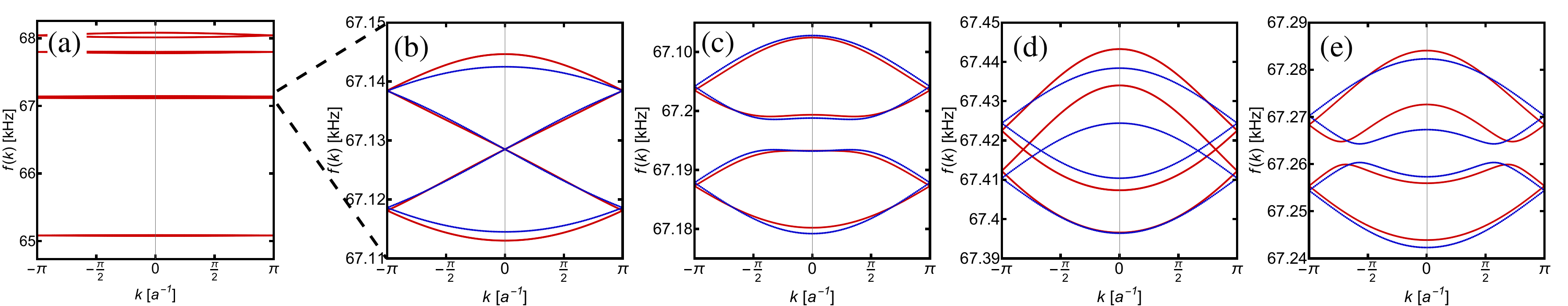}
\caption{ \footnotesize{Band structure of the artificial 1D chain. The red curves correspond to finite-element calculations. The blue curves are obtained from Eq.~\eqref{Hpol}. For the cases considered (in Hz): a) and b) $f_{0} = 67128.5$, $t = 7.0$, and $t' = 0.1$, c) $f_{0} = 67196.0$, $t = 7.2$, and $t' = 3.0$, d) $f_{0} = 67417.4$, $t = 5.0$, and $t' = 5.0$, and e) $f_{0} = 67262.3$, $t = 7.5$, and $t' = 5.0$. } }
\label{fig:BSComsol}
\end{figure*}

{\it Artificial elastic 1D chain with first- and third-nearest-neighbor hoppings.} 
The tight-binding model given in Eq.~\eqref{Hpol} can be implemented with the so-called coupled-resonator elastic waveguides. 
Opposite to other artificial realizations of the tight-binding model with dielectric resonators~\cite{YarivEtAl1999,BayindirEtAl2000,BittnerEtAl2010,Franco-VillafaneEtAl2013,DietzRichter2018}, optical waveguides~\cite{SepkhanovEtAl2007,Longhi2010,Perez-LeijaEtAl2013,SetareEtAl2019,MajariEtAl2021} and defects inside photonic~\cite{YarivEtAl1999} and phononic crystals~\cite{Sigalas,EscalanteMartinezLaude,ReyesWalkerZubovHeoKrokhinNeogi,ReyesEtAl,WangEtAl,Lopez-ToledoEtAl2021}, the coupled-resonator elastic realization offers a high control over the first- and the third-nearest-neighbor hoppings. 
The latter is a very convenient fact for the purpose of the present model. 

The elastic tight-binding implementation is in line with the development of artificial elastic ethylene, butadiene, hexatriene~\cite{RamirezRamirez2020}, benzene~\cite{Martinez-ArguelloEtAl2022}, and borazine~\cite{Mendez-SanchezEtAl2021}, which also satisfy the tight-binding model. As shown in Fig.~\ref{fig:CREW} the structure consists of resonators coupled through finite phononic crystals (FPhCs). When the normal-mode frequencies of the resonators fall within the local bandgap~\cite{RomeroGarcia2013} of the FPhCs, the wave amplitudes localize in the resonators and decay evanescently through the FPhCs; the coupling between localized waves in neighboring resonators is due to evanescent Bloch waves. It is assumed that the FPhCs are large enough that the resonators are weakly coupled to each other. In this regime, the modes in the resonators present a high $Q$ value. The coupling is taken into account to model the transmission of elastic waves and it is the elastic analog of the tight-binding approximation in condensed-matter physics~\cite{Lopez-ToledoEtAl2021}. The Hamiltonian that rules the artificial elastic 1D chain in Fig.~\ref{fig:CREW} is the same as that given in Eq.~\eqref{Hpol} replacing the site energy $E_0$ by the site frequency $f_0$. The couplings $t$ and $t'$ are taken as the evanescent Bragg overlap of the localized modes in different resonators~\cite{Lopez-ToledoEtAl2021}.

{\it Topological phase transition in artificial 1D elastic chains.} 
The topological phase transition, described by the tight-binding model of Eq.~(\ref{Hpol}), can be verified in the corresponding analogous artificial system by using the engineered elastic 2D structure shown in Fig.~\ref{fig:CREW}. The band structure of the artificial chain is shown in Fig.~\ref{fig:BSComsol}, in red curves, for different parameters of the third-nearest-neighbor coupling. 
These band structures are obtained by taking the Floquet boundary conditions on the unit cell with the geometrical parameters given in the caption in Fig.~\ref{fig:CREW} with a) and b) $c_{3} =$5~mm, c) 7~mm, d) 9~mm, and e) 10~mm. For the calculations finite-element simulations, using the software COMSOL Multiphysics for aluminum 1145 with Young's modulus of 70 GPa, density of 2700 $\textrm{Kg}/\textrm{m}^{3}$, and Poisson ratio of 0.33, are considered. The blue curves correspond to the band structure obtained from the tight-binding Hamiltonian in Eq.~\eqref{Hpol}, where the site energy $E_{0}$ and hopping parameters to first and third nearest neighbors are replaced by the site frequency $f_{0}$ and hopping frequencies to first and third nearest neighbors, respectively. 

In Figs.~\ref{fig:BSComsol} b) and c), an excellent agreement between the band structures from the artificial chain and from the tight-binding Hamiltonian in Eq.~\eqref{Hpol} is observed. 
This is not the case for the corresponding ones in panels d) and e), where only good agreement is found. Deviations from the tight-binding model appear: the bands are not symmetric with respect to the frequency axis and the values of $k$ in which the Dirac points appear in case d) are different from those of the tight-binding model. Two possible reasons are 1) the emergent band is not symmetrically centered in the gap of the FPhCs (See Fig.~\ref{fig:BSComsol} a)) and 2) the non-orthogonality of the meta-atom orbitals. In any case, the topological phases predicted by the model of Eq.~(\ref{Hpol}) are fully captured by the elastic chain.

The topological phase transition predicted by Eq.~(\ref{Hpol}) can be tested with the help of the analogous elastic model. This elastic chain can be constructed on an aluminum plate with a large number of unit cells. The measurements can be performed by exciting on one point of a free end of the chain and detecting on the other end using a vector network analyzer (VNA) and the experimental setup used elsewhere~\cite{Martinez-ArguelloEtAl2022}.

In summary, the present effective model is not a simple extension of the SSH model, since it offers the possibility to get two topological phases with opposite winding numbers. Also, a valley-pseudo-spin structure is observed. Finite element simulations of elastic chains indicate that one-dimensional phononic crystals are ideal platforms to test the topological phase transitions (see Figs. \ref{fig:Polyacetylene} and \ref{fig:BSComsol}). Although the modified Dirac Hamiltonian in Eq. \eqref{Heff} was developed at the nanoscale for electrons in cis-polyacetylene with first and third nearest neighbors only, the twofold topological phase was compared at the macroscopic scale for vibrations in elastic chains based on aluminium. This fact shows the universality phenomena of this topological phase transition regardless of the scale. The latter is a consequence of the omnipresence of the Bloch theorem, both at the microscopic and macroscopic levels, for periodic arrangements.

{\it Acknowledgments.} Y.B.-O. and R.A.M.-S. acknowledge financial support from UNAM-PAPIIT under projects IA106223 and IN111021, respectively. Y.B.-O. was partially supported by DGAPA-UNAM through the POSDOC program. A.M.M.-A. and B.M.-M. acknowledge financial support from CONHACyT. 


\end{document}